# Orbit-transfer torque driven field-free switching of perpendicular magnetization

Xing-Guo Ye[†], Peng-Fei Zhu[†], Wen-Zheng Xu[†], Nianze Shang, Kaihui Liu, Zhi-Min Liao*

State Key Laboratory for Mesoscopic Physics and Frontiers Science Center for Nano-optoelectronics, School of Physics, Peking University; Beijing 100871, China.

*Corresponding author. Email: liaozm@pku.edu.cn

†These authors contributed equally to this work.

The reversal of perpendicular magnetization (PM) by electric control is crucial for high-density integration of low-power magnetic random-access memory (MRAM). Although the spin-transfer torque (STT) and spin-orbit torque (SOT) technologies have been used to switch the magnetization of a free layer with perpendicular magnetic anisotropy, the former has limited endurance because of the high current density directly through the junction, while the latter requires an external magnetic field or unconventional configuration to break the symmetry. Here we propose and realize the orbit-transfer torque (OTT), that is, exerting torque on the magnetization using the orbital magnetic moments, and thus demonstrate a new strategy for current-driven PM reversal without external magnetic field. The perpendicular polarization of orbital magnetic moments is generated by a direct current in a few-layer WTe$_2$ due to the existence of nonzero Berry curvature dipole, and the polarization direction can be switched by changing the current polarity. Guided by this principle, we construct the WTe$_2$/Fe$_3$GeTe$_2$ heterostructures, where the OTT driven field-free deterministic switching of PM is achieved.

**Main Text:** MRAM[1-4] as a type of non-volatile memory has advantages of high speed, low power consumption and high reliability, which is recently being developed by using magnetic tunnel junction (MTJ) with perpendicular magnetic anisotropy (PMA) for further scaling down[5]. The PM reversal of the free layer is the key operation of data writing in MRAM[6]. The magnetization switch driven by STT effect[7-10] has been implemented in two-terminal MTJ, which challenges its durability due to the high-density current passing through the junction. Thereafter, three-terminal SOT[11-28] devices have been developed, where the magnetization switching is caused by interfacial spin accumulations coming from spin Hall effect[13,14], Rashba effect[12], or the topological surface states[16]. However, an external magnetic field is usually needed for the PM switching driven by SOT effect[15], although a few experiments have demonstrated the SOT-induced filed-free PM switching by introducing interfacial exchange bias[22] or structure asymmetry[17,25], which is unfavorable for realistic applications. Here we propose and demonstrate the OTT effect to obtain the current-driven field-free PM switching. Instead of spin degree of freedom, the OTT effect exploits the polarization of orbital magnetic moment of Bloch electrons[29].

**Orbit-transfer torque as a new strategy for PM switching**

The recently discovered Berry curvature dipole[30] in Van der Waals materials offers an ideal platform to generate the perpendicularly polarized orbital magnetic moments by applying dc current[31], formulated as $\boldsymbol{m} \propto (\boldsymbol{D} \cdot \boldsymbol{E})\hat{\boldsymbol{z}}$, where $\boldsymbol{D}$ is the Berry curvature dipole and $\boldsymbol{E}$ is the applied electric field. Similar with the SOT, OTT also has two components, that is, field-like torque $T_{FL} \sim \boldsymbol{M} \times \hat{\boldsymbol{m}}$, and antidamping-like torque $T_{AD} \sim \boldsymbol{M} \times (\hat{\boldsymbol{m}} \times \boldsymbol{M})$ (Ref.[23]), where $\boldsymbol{M}$ is the magnetization of adjacent magnetic layer, and $\hat{\boldsymbol{m}}$ is the unit vector of polarized orbital magnetic moment. For perpendicularly polarized $\hat{\boldsymbol{m}}$, nonzero antidamping-like torque would emerge as long as $\boldsymbol{M}$ has a slight deviation from the perpendicular direction. For ferromagnetic materials with PMA, either spin wave excitation[32] or spin fluctuations can produce a perturbation of spin orientation deviating from the perpendicular direction. Therefore, the OTT can force the magnetization to prefer along the orbital polarization direction, as shown in **Fig. 1a,b**.

A simplified analysis is carried out to search the candidate materials to generate OTT. First, we focus on the 2D layered materials, where the dimension constraint forces the orbital magnetic moment along out-of-plane direction. Second, for 2D materials with nonzero Berry curvature dipole, such as bilayer[33] or few-layer $WTe_2$ (Ref.[34]), and strained $WSe_2$ (Ref.[35]), current can induce the polarization of orbital magnetic moment[31]. As shown in **Fig. 1c**, Berry curvature distributes asymmetrically in one single valley in these materials, forming a Berry curvature dipole. Applying a parallel dc current, the orbital magnetic moment is thus polarized. Further, as shown in **Fig. 1d**, when the applied current polarity is reversed, the polarization direction of the orbital magnetic moment is also reversed.

Guided by this principle, we constructed the few-layer $WTe_2$/$Fe_3GeTe_2$ heterostructures (see Device Fabrication in Supplemental Material), where the few-layer $WTe_2$ with nonzero Berry curvature dipole is used to generate the polarization of orbital magnetic moment by applying current, and the few-layer $Fe_3GeTe_2$ is used as the ferromagnetic layer with PMA[36]. Field-free current-driven magnetization switching has been demonstrated as a result of the OTT effect at the heterostructure interface. The measurement results of four devices, device A in the main text and devices B-D in the Figs. S7-9 and Table S1 of Supplemental Material, are presented.

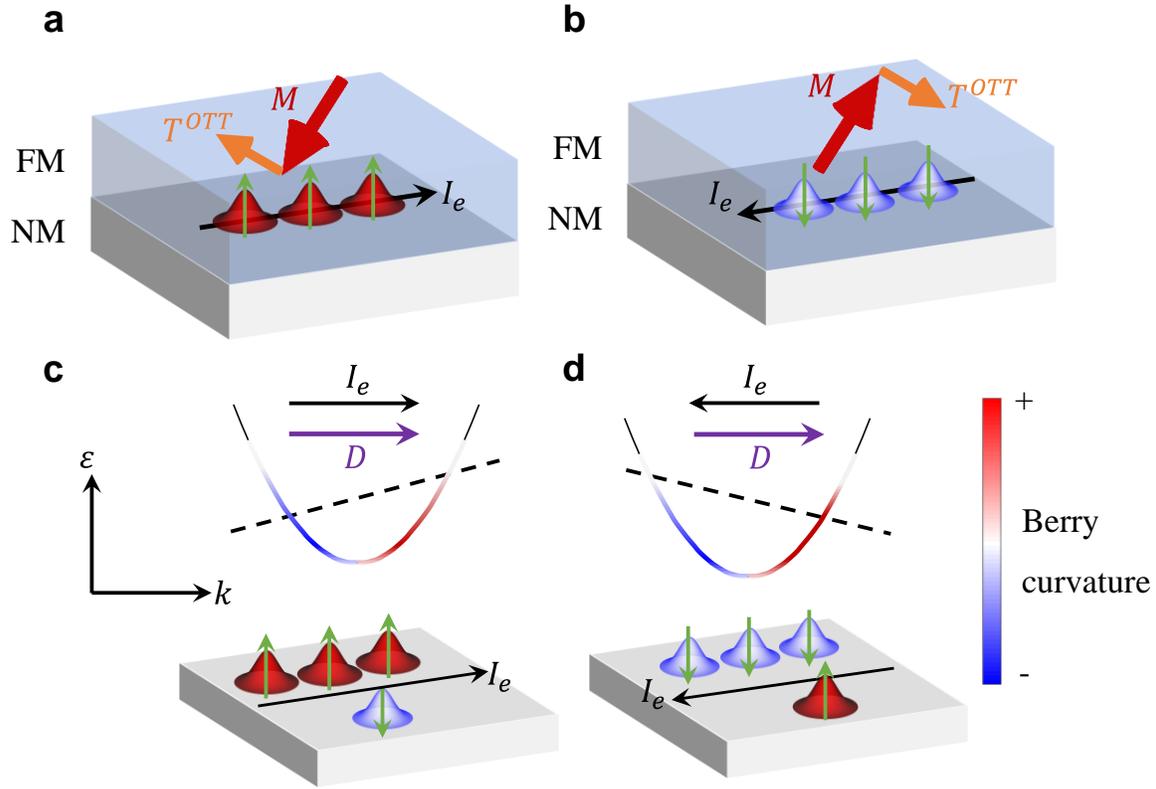

**Fig. 1 | Illustration of orbit-transfer torque. a,b,** For the ferromagnetism (FM)/non-magnetic material (NM) heterostructure, the polarization of orbital magnetic moment can result in orbit-transfer torque that acts on the magnetization, and the up and down magnetization states can be switched by changing the current direction. **c,d,** Illustration of current-induced polarization of orbital magnetic moment in systems with nonzero Berry curvature dipole. The orbital magnetic moment is denoted by green arrow. Bloch electrons with opposite Berry curvature are denoted by red and blue, respectively. $I_e$ is the electron current. $D$ is the Berry curvature dipole. $\varepsilon$ is the energy and $k$ is the wavevector.

## Polarization of orbital magnetic moment manifested by nonlinear Hall effect

WTe$_2$ is a 2D layered van der Waals material belonging to the transition metal dichalcogenide family, with T$_d$-phase as its stable structure[37]. Monolayer T$_d$-WTe$_2$ consists of a layer of W atoms sandwiched between two layers of Te atoms in a distorted octahedral coordination, as shown in **Fig. 2a**. WTe$_2$ possesses various exotic phases, including Weyl semimetal[38], gate-tunable superconductivity[39,40], quantum spin Hall insulator[41,42], and exciton insulator[43], depending on its thickness. Its bulk crystal has space group $Pmn2_1$, which possesses a mirror symmetry along $b$ axis mirror line (red line in Fig. 2a) and a glide mirror symmetry along $a$ axis mirror line[34]. These two mirror symmetries recover an inversion symmetry in the $ab$ plane, leading to vanishing Berry curvature dipole in WTe$_2$ bulk. However, since the glide mirror symmetry involves a half-cell translation, it is generally broken at the surfaces, giving rise to nonzero Berry curvature dipole at surfaces[44]. Because the mirror symmetry with $bc$ plane as mirror plane still exists at the surface, it constrains the Berry curvature dipole in WTe$_2$ along $a$ axis direction. The nonzero Berry curvature dipole in few-layer WTe$_2$ has been observed in previous experiments[34], evidenced by the nonlinear Hall effect.

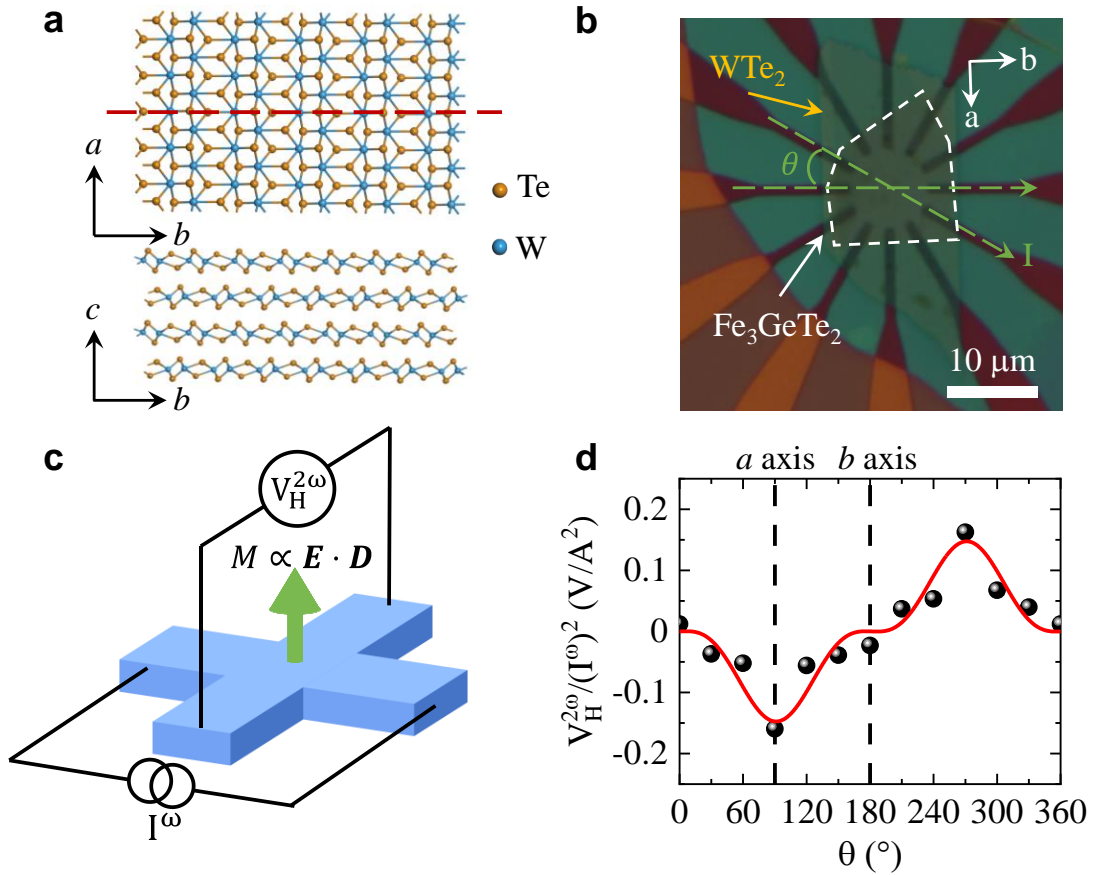

**Fig. 2 | Nonlinear Hall effect in WTe₂/Fe₃GeTe₂ heterostructures. a,** Crystal structure of T$_d$-WTe$_2$. The red line corresponds to the mirror line in the $ab$ plane, which persists at the surface. **b,** The optical image of device A, which consists of a WTe$_2$/Fe$_3$GeTe$_2$ heterostructure. The crystalline

axes are aligned with electrodes. The $\theta$ is defined as the angle between the current direction and the baseline direction of the electrode pair, which is approximately along the $b$ axis of $WTe_2$. For device A, the baseline direction is ~1.3° misaligned with $b$ axis. **c,** Schematic of the measurement configuration for nonlinear Hall effect. **d,** The angle-dependence of the second-order nonlinear Hall effect at 1.8 K. The red line is the fitting curve.

The optical image of fabricated $WTe_2/Fe_3GeTe_2$ heterostructure, device A, is shown in Fig. 2b. The thickness of $WTe_2$ and $Fe_3GeTe_2$ measured by atomic force microscope is 11.9 and 11.2 nm, respectively (see Fig. S1). The multiple electrodes were designed as a circular disc configuration, which was utilized to implement the angle-dependent measurements. By identifying long, straight edges of $WTe_2$ and combining with polarized Raman spectroscopy, the crystalline axes were aligned with the electrodes, where the misalignment is ~1.3° for device A (see Fig. S2). The angle $\theta$ is defined in Fig. 2b, where $\theta = 0°$ approximately corresponds to $b$ axis; while $\theta = 90°$ approximately corresponds to $a$ axis. The device shows two carrier transport properties and large magnetoresistance with electron mobility ~2203 $cm^2/V \cdot s$ (see Fig. S3). Since nonzero Berry curvature dipole emerges at the interface, when applying an electric field, it would give rise to the polarization of orbital magnetic moment, expressed as $\boldsymbol{m} \propto (\boldsymbol{D} \cdot \boldsymbol{E})\hat{\boldsymbol{z}}$ (Ref.[31,45]). When an a.c. current $I^\omega$ with frequency $\omega$ is applied, the induced orbital magnetization will also have the alternating characteristic of frequency $\omega$, further resulting in a second-harmonic anomalous Hall voltage $V_H^{2\omega}$ with frequency $2\omega$, known as the nonlinear Hall effect[35], as illustrated in Fig. 2c. The results are displayed in Fig. S4a, and quadric $I^\omega - V_H^{2\omega}$ characteristics are observed. Moreover, the nonlinear Hall voltage shows maximum at $\theta = 90°$ approximately along $a$ axis and is nearly zero at $\theta = 0°$ approximately along $b$ axis (see Fig. 2d). This angle-dependence is consistent with the fact that the Berry curvature dipole in few-layer $WTe_2$ is along the crystal $a$ axis.

**Orbit-transfer torque in WTe₂/Fe₃GeTe₂ heterostructures**

The hysteresis loops of Hall resistance ($R_{xy}$) of the WTe$_2$/Fe$_3$GeTe$_2$ heterostructure were measured by sweeping the out-of-plane magnetic field at various temperatures. From the Arrott plots (see Fig. S5), it is found that the Curie temperature is ~180 K, corresponding to that of a 14-layer Fe$_3$GeTe$_2$ based on previous work[36]. As shown in Fig. 3a and Fig. S6, the $R_{xy}$ hysteresis loops with sharp transitions are clearly observed, consistent with the PMA characteristics of Fe$_3$GeTe$_2$ (Ref.[36,46]). To demonstrate the field-free switching of magnetization in Fe$_3$GeTe$_2$ layer through OTT, a pulse-like d.c. current I$_P$ was injected into the heterostructure at $\theta = 90°$ approximately along *a* axis, and the $R_{xy}$ was measured after the pulse I$_p$ was removed. As shown in Fig. 3b, current-induced $R_{xy}$ change is clearly observed at 130 K. The height of the $R_{xy}$-I$_P$ loop agrees well with that of the $R_{xy}$-H loop in Fig. 3a, indicating the deterministic PM switching. Therefore, the $R_{xy}$-I$_P$ loops indicate that the upward and downward magnetization states in Fe$_3$GeTe$_2$ can be switched between each other by injecting opposite currents. Since no external magnetic field is applied, such PM switching is failed to be attributed to the SOT effect in WTe$_2$ (Ref.[27]). The current induced PM switching agrees well with the OTT effect, that is, the torque exerted by the perpendicularly polarized orbital magnetic moments.

A sequence of current pulse with amplitude $\pm 8$ mA at 130 K and $\theta = 90°$ was injected into this heterostructure, as shown in Fig. 3c. Current pulse with opposite polarity yields terminate states with saturated magnetization along opposite directions (up or down). Once the magnetization state is finalized, the disturbing current pulse will not affect the magnetization state, as indicated by the red arrows in Fig. 3c. Such robust magnetization switching against perturbations indicates the application potential in MRAM. Temperature dependence of this current-induced PM switching is further investigated, as shown in Fig. 3d. Two critical switching current densities J$_c$ and J$_{on}$ are defined, where J$_c$ corresponds to the current density at which resistance changes by more than half, and $J_{on}$ corresponds to the current density at the onset of resistance change. Both $J_c$ and $J_{on}$ are taken as the average for up-to-down and down-to-up PM switching. As shown in Fig. 3e, the $J_c$ and $J_{on}$ decrease upon increasing temperature, consistent with the fact that the effective PMA field in $Fe_3GeTe_2$ decreases upon increasing temperature.

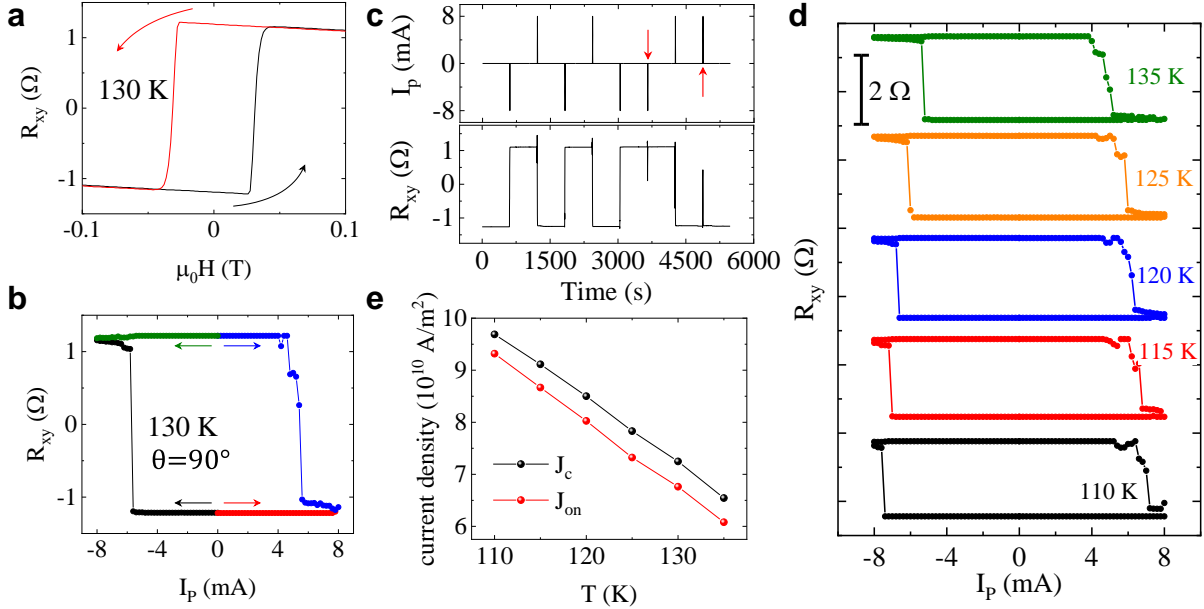

**Fig. 3 | Orbit-transfer torque induced field-free perpendicular magnetization switching in WTe₂/Fe₃GeTe₂ heterostructure. A,** Hall resistance as a function of magnetic field at 130 K. The hysteresis loop is used for reference to identify the magnetization states. **b,** Hall resistance as a function of pulse current $I_P$ at 130 K. Before sweeping $I_P$ from zero to large positive or negative values, the magnetization state is initialized by magnetic field. **C,** Top panel: A series of current pulse with amplitude ±8 mA is injected at 130 K. The red arrows indicate the deliberately programmed disturbing current pulses. Bottom panel: The corresponding changes in Hall resistance. **d,** Temperature dependence of the current-driven magnetization switching. The curves are shifted for clarity. **e,** The temperature dependence of critical switching current density. The $J_c$ corresponds to the current density at which resistance changes by more than half, and $J_{on}$ corresponds to the current density at the onset of resistance change. The data in this figure were measured when the current was applied approximately along $a$ axis at $\theta = 90°$.

**Synergistic effect of spin-orbit torque and orbit-transfer torque**

The above results indicate the OTT-driven PM switching as the current was applied along *a* axis of WTe$_2$. It would be very different as the pulse current I$_P$ is applied at $\theta = 0°$ approximately along *b* axis, where the current direction is perpendicular to the Berry curvature dipole and there is almost no current induced orbital magnetism. As shown in Fig. 4a, no deterministic switching is observed when I$_P$ is along *b* axis at 130 K. Instead of switching between upward and downward PM states, I$_P$ along *b* axis yields a terminate state with nearly zero Hall resistance. This is well understood by the SOT-induced multi-domain scenario[28]. Since SOT forces the magnetization lying in plane, when removing pulse current and thus SOT, multi-domains with magnetization randomly upward or downward would emerge, leading to nearly zero Hall resistance. Thus, the OTT is absent, and SOT dominates when I$_P$ is along *b* axis, as expected.

Using the disc distributed electrode structure of the device, angle-dependence of current-driven magnetization switching was investigated at 130 K, as shown in Fig. 4b. It is found that the deterministic switching can be obtained within a large angle window due to the nonzero current component along *a* axis, as shown by the R$_{xy}$-I$_P$ loops at $\theta = 120°$ and 60° in Fig. 4b. As for the current along direction closer to *b* axis, that is, $\theta = 30°$ and 150°, the R$_{xy}$-I$_P$ loops only shows partially switching without reaching the saturated magnetization states, which suggests the coexistence of SOT and OTT. The angle-dependent critical switching current density is shown in Fig. 4c. The J$_c$ is failed to be fitted in the form of $1/cos(\theta - 90°)$ as only considering the current component parallel to *a* aixs. Because the bulk states can also contribute to the SOT due to intrinsic spin-orbit coupling and the OTT only comes from the interface due to the need of symmetry breaking, the strength of SOT should be larger than that of OTT in the WTe$_2$/Fe$_3$GeTe$_2$ device. Therefore, the SOT may dominate the onset change of magnetic states, but the deterministic switching is achieved through OTT. The angle dependence of the onset switching current density J$_{on}$ is consistent with the anisotropy of spin-orbit coupling in bulk WTe$_2$, which possesses the largest strength along $b$ axis[27]. It is worth noting that the critical switching current at $\theta = 60°$ and $120°$ shows a smaller value than that at $\theta = 90°$, suggesting that the synergy of SOT and OTT is more effective for the PM switching.

It is worth noting that some works point out that out-of-plane spin accumulation may occur together with SOT in $WTe_2$ (ref. 47). Although this out-of-plane spin accumulation might be a possible origination of the field-free perpendicular magnetization switching, it can be ruled out in our work due to the following two reasons. First, our results demonstrate the anisotropic angle dependence of the magnetization switching, which is not expected for the mechanism of out-of-plane spin accumulation. In contrast, this angle dependence is naturally explained by OTT. Second, the SOT in $WTe_2$ was reported to be mainly contributed by bulk states[27], however, our results show that the switching critical current density is almost the same in samples with different thickness (see Table S1), indicating a surface origin of the antidamping torque. This surface origin is also well consistent with the mechanism of OTT.

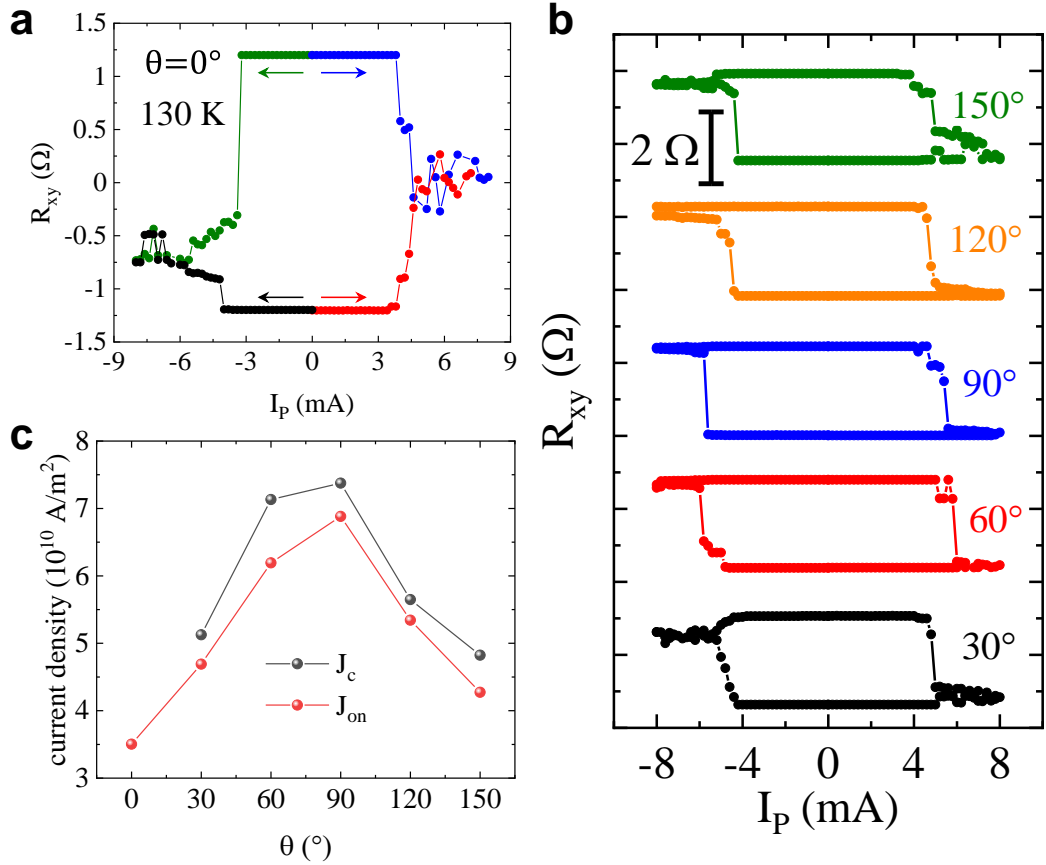

**Fig. 4 │ Angle-dependence of current-driven magnetization switching in WTe₂/Fe₃GeTe₂ heterostructure.** **a,** Hall resistance as a function of pulse current $I_P$ at 130 K. $I_P$ is applied at $\theta = 0°$ approximately along $b$ axis. Before sweeping $I_P$ from zero to large positive or negative values, the magnetization state is initialized by magnetic field. **b,** The $R_{xy}$-$I_P$ loops at 130 K with $I_P$ applied along different angle $\theta$. The curves are shifted for clarity. **c,** The critical switching current density as a function of angle $\theta$.

**Outlook**

Our work demonstrates the OTT as a new strategy for field-free magnetization switching in PMA ferromagnets, which utilizes the orbital magnetic moment of Bloch electrons rather than spins, thus extending the concept from spintronics to orbitronics and their hybrid. Through symmetry analysis, we propose that besides few-layer WTe₂, 2D materials with nonzero Berry curvature dipole[48], including bilayer WTe₂ (Ref.[33]), strained WSe₂ (Ref.[35]), few-layer MoTe₂ (Ref.[49]), and corrugated bilayer graphene[50], are candidates to realize OTT. Moreover, the temperature dependence of current-driven switching indicates OTT is robust against temperature. By building heterostructures using PMA materials with Curie temperature above 300 K, such as few-layer $CrTe_2$ (Ref.[51]), room-temperature PM switching through OTT can be achieved, promising for realistic applications.

**Acknowledgments:** This work was supported by National Natural Science Foundation of China (Grant Nos. 91964201 and 61825401).

**Author contributions:** Z.M. L. conceived and supervised the project. P.F.Z., X.G.Y., and W.Z.X. fabricated the devices and performed the transport measurements. N.S., X.G.Y. and K.L. performed the polarized Raman spectroscopy measurement. Z.M. L. and X.G.Y. analyzed the data and wrote the manuscript with discussion and input of all authors.

**Supplemental Material for**

**Orbit-transfer torque driven field-free switching of perpendicular magnetization**

Xing-Guo Ye[†], Peng-Fei Zhu[†], Wen-Zheng Xu[†], Nianze Shang, Kaihui Liu, Zhi-Min Liao*

State Key Laboratory for Mesoscopic Physics and Frontiers Science Center for Nano-optoelectronics, School of Physics, Peking University; Beijing 100871, China.

*Corresponding author. Email: liaozm@pku.edu.cn

†These authors contributed equally to this work.

**Device Fabrication**

Few layer $WTe_2$ was obtained from high-quality artificially grown crystals of bulk $WTe_2$ commercially purchased from HQ Graphene through standard mechanically exfoliated method. Few layer $Fe_3GeTe_2$ was obtained in a similar way. Then we patterned Ti/Au electrodes (~10 nm thick) onto an individual $SiO_2$/Si substrate with a circular disc configuration through e-beam lithography, metal deposition and lift-off. To achieve a better contact, the electrodes were precleaned by air plasma. Exfoliated BN flake (~20 nm thick), few layer $Fe_3GeTe_2$ (~10-15 nm thick) and few layer $WTe_2$ (~2-15 nm thick) were sequentially picked up and then transferred onto the Ti/Au electrodes using a polymer-based dry transfer technique[52]. The whole exfoliated and transfer processes were done in an argon-filled glove box with $O_2$ and $H_2O$ content below 0.01 parts per million to avoid sample degradation.

**Polarized Raman spectroscopy of few-layer $WTe_2$**

The Raman spectroscopy was measured with 514 nm excitation wavelengths through a linearly polarized solid-state laser beam. The polarization of the excitation laser was controlled by a half-wave plate and a polarizer. The Raman scattered light with the same polarization as the excitation laser were collected. As shown in Fig. S2b, five Raman peaks are observed, which belong to the A1 modes of $WTe_2$ (Ref.[53]). The polarization dependence of intensities of peaks P2 and P11 (denoted in Fig. S2b) are presented in Figs. S2c and S2d, respectively. Based on previous reports[53], the polarization direction with maximum intensity was assigned as the $b$ axis. The determined crystalline axes, i.e., $a$ axis and $b$ axis, are further denoted by the black arrows in the optical image (Fig. S2a).

**Transport Measurements**

All transport measurements were carried out in an Oxford cryostat with a variable temperature insert and a superconducting magnet. First-, second- and third-harmonic voltage signals were collected by standard lock-in techniques (Stanford Research Systems Model SR830) with frequency $\omega$ = 17.777 Hz unless otherwise stated. A sequence of pulse-like d.c. current $I_P$ was applied through a Keithley 2400 SourceMeter. $I_P$ was swept in steps of 0.2 mA. After every $I_P$ was applied and then removed, the Hall resistance was measured as applying a 0.1 mA bias a.c. current.

**Basic Transport Properties of Device A**

The resistivity $\rho_{xx}$ of device A along $a$ axis as a function of temperature was measured, as shown in Fig. S3a. By considering a parallel resistance model, we could obtain the resistivity of WTe$_2$ by $\rho_{xx}^{WTe_2} = t_{WTe_2}\Big/\left(\frac{t}{\rho_{xx}} - \frac{t_{FGT}}{\rho_{xx}^{FGT}}\right)$, where $\rho_{xx}^{WTe_2}$ is the resistivity of WTe$_2$, $\rho_{xx}^{FGT}$ is the resistivity of Fe$_3$GeTe$_2$, $t_{WTe_2}$ is the thickness of WTe$_2$, $t_{FGT}$ is the thickness of Fe$_3$GeTe$_2$ and $t = t_{WTe_2} + t_{FGT}$. Referring to the $\rho_{xx}^{FGT}(T)$ previously reported by Z. Fei *et al* (Ref.[54]), the $\rho_{xx}^{WTe_2}$ was calculated and presented in Fig. S3a. Furthermore, the fraction of current flowing in the WTe$_2$ layer is estimated by $\frac{I_{WTe_2}}{I} = \frac{1}{1+\frac{\rho_{xx}^{WTe_2}}{\rho_{xx}^{FGT}}\frac{t_{FGT}}{t_{WTe_2}}}$, where $I$ is the applied current flowing in the whole heterostructure, and $I_{WTe_2}$ is the current component flowing in the WTe$_2$ layer. For other angles, $\rho_{xx}^{WTe_2}$ was estimated through the intrinsic resistivity anisotropy of WTe$_2$ following $\rho_{xx}(\theta) =$ $\rho_a sin^2(\theta - \theta_0) + \rho_b cos^2(\theta - \theta_0)$, where $\rho_a$ and $\rho_b$ are resistivity along $a$ axis and $b$ axis, respectively, $\theta_0$ corresponds to $b$ axis.

The magneto-transport properties of device A at 1.8 K are shown in Fig. S3b. The large non-saturated magnetoresistance and Hall resistance demonstrate two-carrier transport characteristics, indicating a nearly compensated electron and hole density in WTe$_2$. Through a semi-classical two-carrier model[55], that is, $\rho_{xx} = \frac{1}{e} \frac{n\mu_n + p\mu_p + (n\mu_p + p\mu_n)\mu_n\mu_p B^2}{(n\mu_n + p\mu_p)^2 + (n-p)^2 \mu_n^2 \mu_p^2 B^2}$ and $\rho_{xy} = \frac{1}{e} \frac{(p\mu_p^2 - n\mu_n^2)B + (p-n)\mu_n^2\mu_p^2 B^3}{(n\mu_n + p\mu_p)^2 + (n-p)^2 \mu_n^2 \mu_p^2 B^2}$, where $n$ is the electron density, $p$ is the hole density, $\mu_n$ is the electron mobility and $\mu_p$ is the hole mobility, the carrier density and mobility are estimated as, $n = 1.51 \times 10^{13}\ cm^{-2}$, $p = 0.99 \times 10^{13}\ cm^{-2}$, $\mu_n = 2203\ cm^2/V \cdot s$ and $\mu_p = 1497\ cm^2/V \cdot s$.

**Higher-order Hall effect in WTe$_2$**

Due to the nonzero Berry curvature dipole on the surface, second-order nonlinear Hall effect is expected in few-layer WTe$_2$ (Ref.[34]). By utilizing the disc geometry of the electrodes, angle-dependence of the second-order nonlinear Hall effect was investigated, as shown in Fig. 2, which could help to confirm the alignment between electrodes and crystalline axis of WTe$_2$. Based on the symmetry of WTe$_2$, the second-order nonlinear Hall effect shows angle-dependence following $\frac{V_H^{2\omega}}{(I^\omega)^2} \propto sin(\theta - \theta_0)[d_{12}r^2 sin^2(\theta - \theta_0) + (d_{11} - 2d_{26}r^2)cos^2(\theta - \theta_0)]$ (Ref.[30]), where $V_H^{2\omega}$ is the second-harmonic Hall voltage, $I^\omega$ is the applied a.c. current, $r$ is the resistance anisotropy, $d_{ij}$ are the elements of the second-order nonlinear susceptibility tensor for the $Pm$ point group, $\theta_0$ is the angle misalignment between $\theta = 0°$ and crystalline $b$ axis. The fitting curve for this angle dependence is shown by the red line in Fig. 2d, which yields the misalignment $\theta_0$ equals 1.3° in device A.

In addition to the second-order nonlinear Hall effect, it is recently also reported a third-order nonlinear Hall effect in the bulk of WTe$_2$ induced by the Berry connection polarizability tensor[56].

Fig. S4 shows the third-order nonlinear Hall effect in device A at 1.8 K. The third-order nonlinear Hall effect shows angle-dependence following $\frac{V_H^{3\omega}}{(I^\omega)^3} \propto cos(\theta - \theta_0)sin(\theta - \theta_0)[(\chi_{22}r^4 - 3\chi_{12}r^2)sin^2(\theta - \theta_0) + (3\chi_{21}r^2 - \chi_{11})cos^2(\theta - \theta_0)]$ (Ref.[47]), where $V_H^{3\omega}$ is the third-harmonic Hall voltage, $\chi_{ij}$ are elements of the third-order susceptibility tensor. The fitting curve for this angle dependence is shown by the red line in Fig. S4c, which yields a similar misalignment angle $\theta_0 \sim 1.3°$.

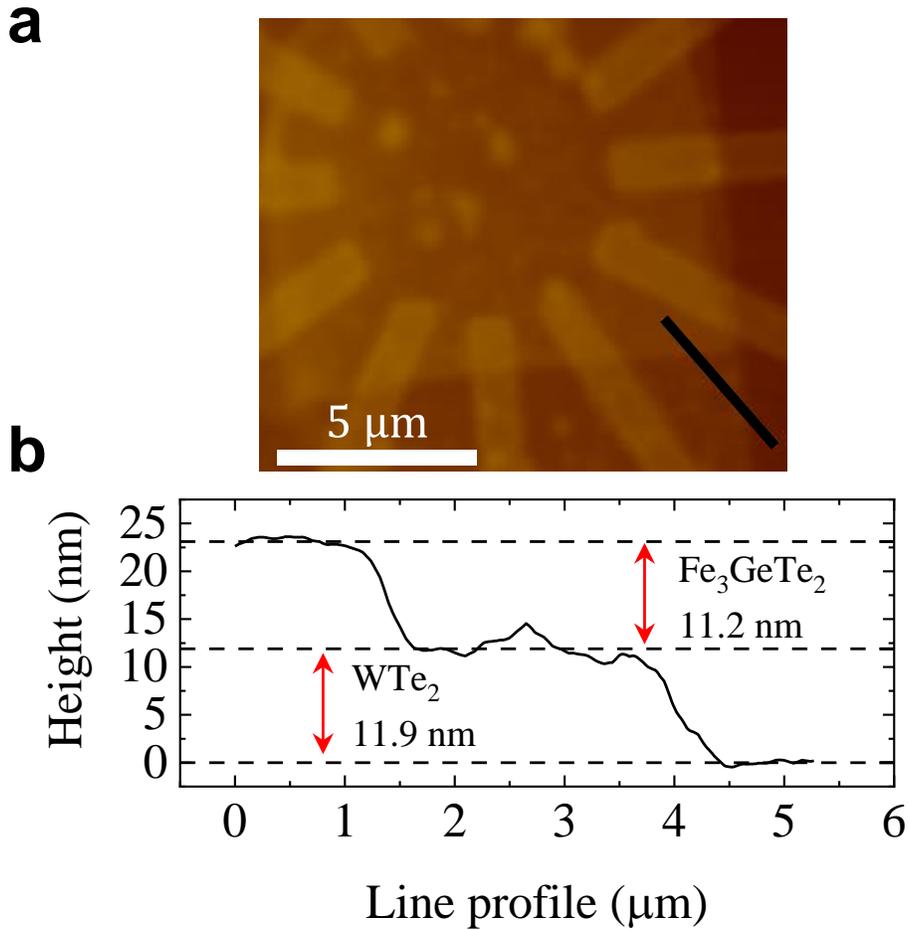

**Fig. S1 | The atomic force microscope image of device A.** The line profile shows the thickness of the $WTe_2$ is 11.9 nm, corresponding to 17-layer thickness. The thickness of the $Fe_3GeTe_2$ is 11.2 nm, corresponding to 14-layer thickness.

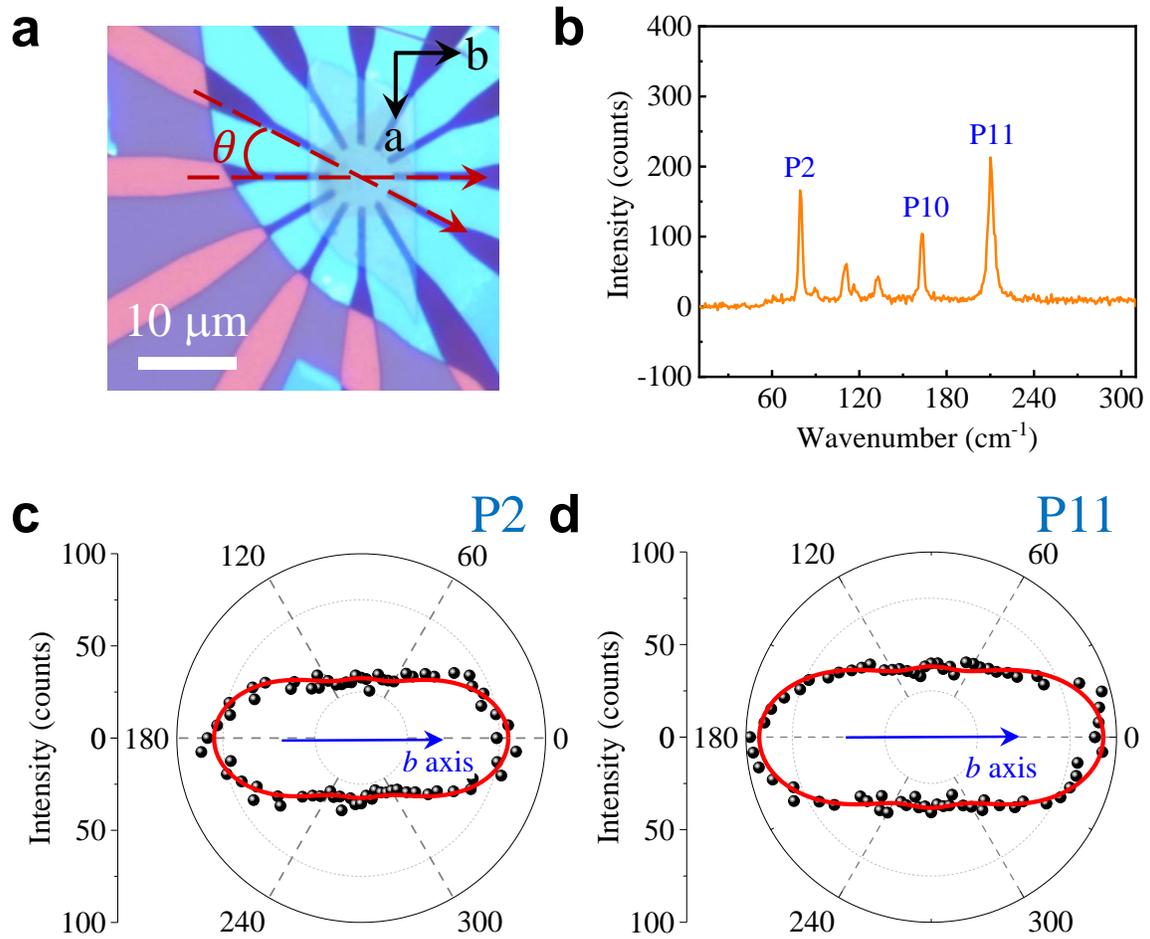

**Fig. S2 | Polarized Raman spectroscopy of few-layer WTe₂ to determine the crystalline orientation. a,** The optical image of device A. **b,** A typical Raman spectrum of device A, where the polarization direction is approximately along *b* axis. **c,d,** Polarization dependence of intensities of peaks (**c**) P2 and (**d**) P11 for device A.

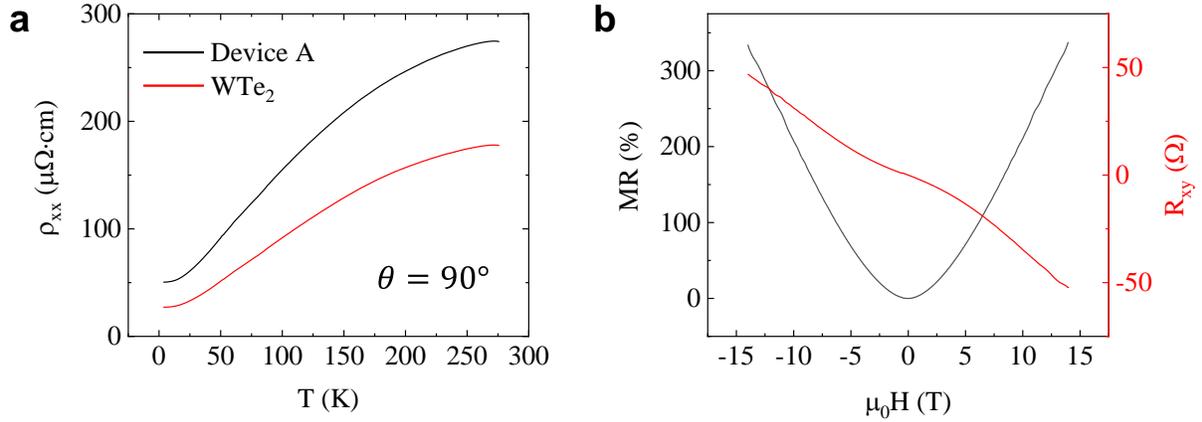

**Fig. S3 │ Basic transport properties of device A. a,** The resistivity as a function of temperature. **b,** Magnetoresistance (MR) and Hall resistance as a function of magnetic field at 1.8 K, marked by black and red, respectively. The MR is defined as $\frac{R_{xx}(\mu_0 H) - R_{xx}(0)}{R_{xx}(0)} \times 100\%$. The large non-saturated MR and Hall resistance demonstrate two-carrier transport characteristics, indicating a nearly compensated electron and hole density in WTe$_2$.

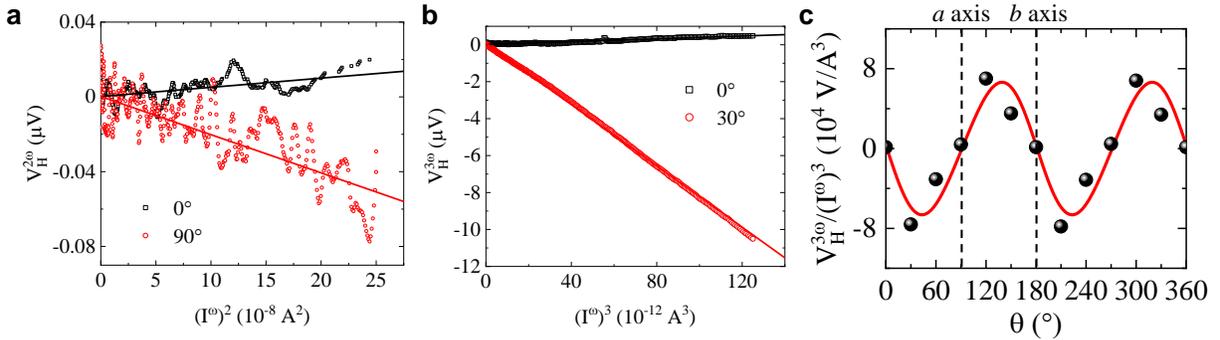

**Fig. S4 │ Higher-order nonlinear Hall effect in device A at 1.8 K. a,** The second-order nonlinear Hall voltage as a function of $(I^\omega)^2$, where $I^\omega$ along $\theta = 0°$ and 90° is marked by black and red, respectively. **b,** The third-order nonlinear Hall voltage as a function of $(I^\omega)^3$, where $I^\omega$ along $\theta = 0°$ and 30° is marked by black and red, respectively. **c,** The angle dependence of the third-order nonlinear Hall effect.

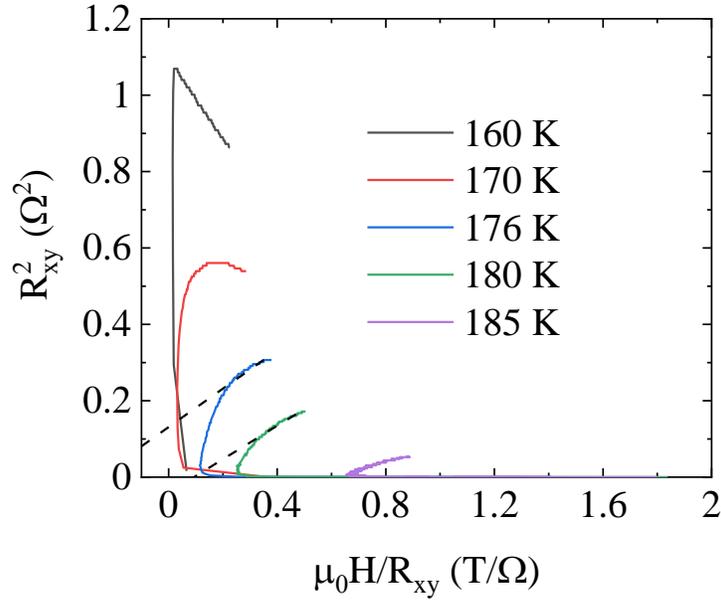

**Fig. S5 │ Arrott plot of device A.** The Curie temperature ~180 K is estimated.

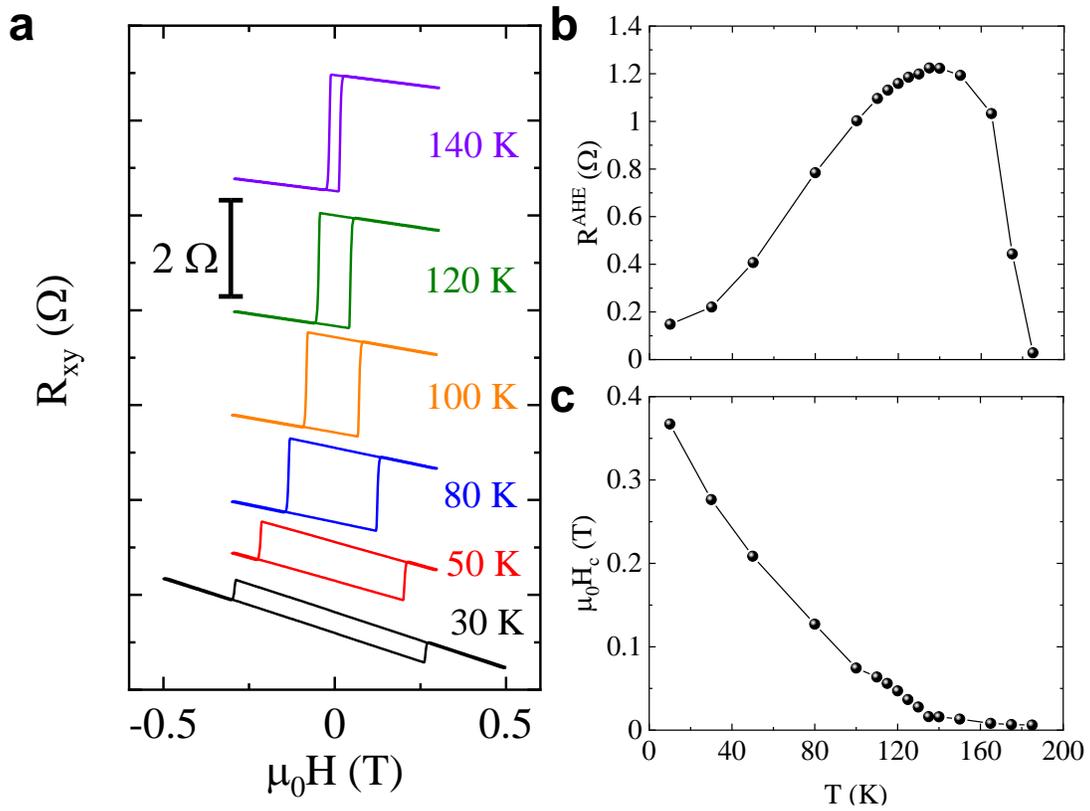

**Fig. S6 │ Magnetic properties of device A. a,** The Hall resistance as a function of magnetic field at various temperatures. **b,** The anomalous Hall resistance, defined as the half of the R-H loop height, as a function of temperature. **c,** The coercive field as a function of temperature.

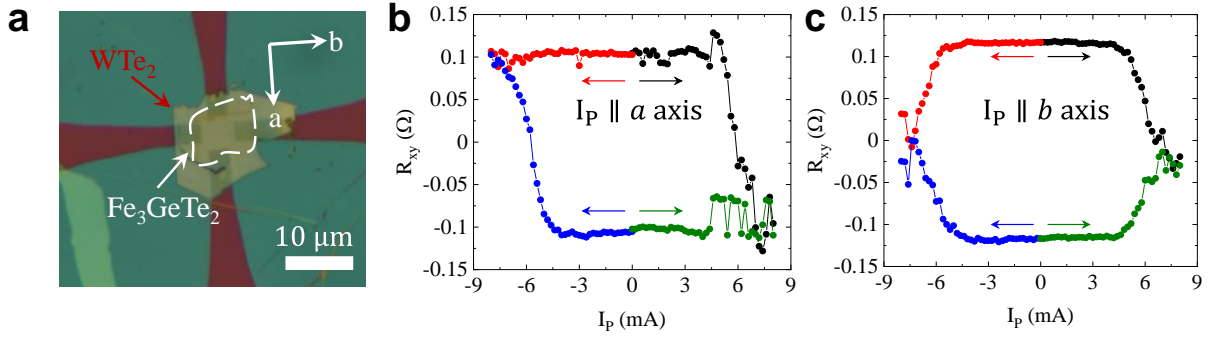

**Fig. S7 │ Reproducible results in device B. a,** The optical image of device B. **b, c,** Hall resistance as a function of pulse current $I_P$ at 120 K for $I_P$ approximately along $a$ axis and $b$ axis, respectively. Before sweeping $I_P$ from zero to large positive or negative values, the magnetization state is initialized by perpendicular magnetic field.

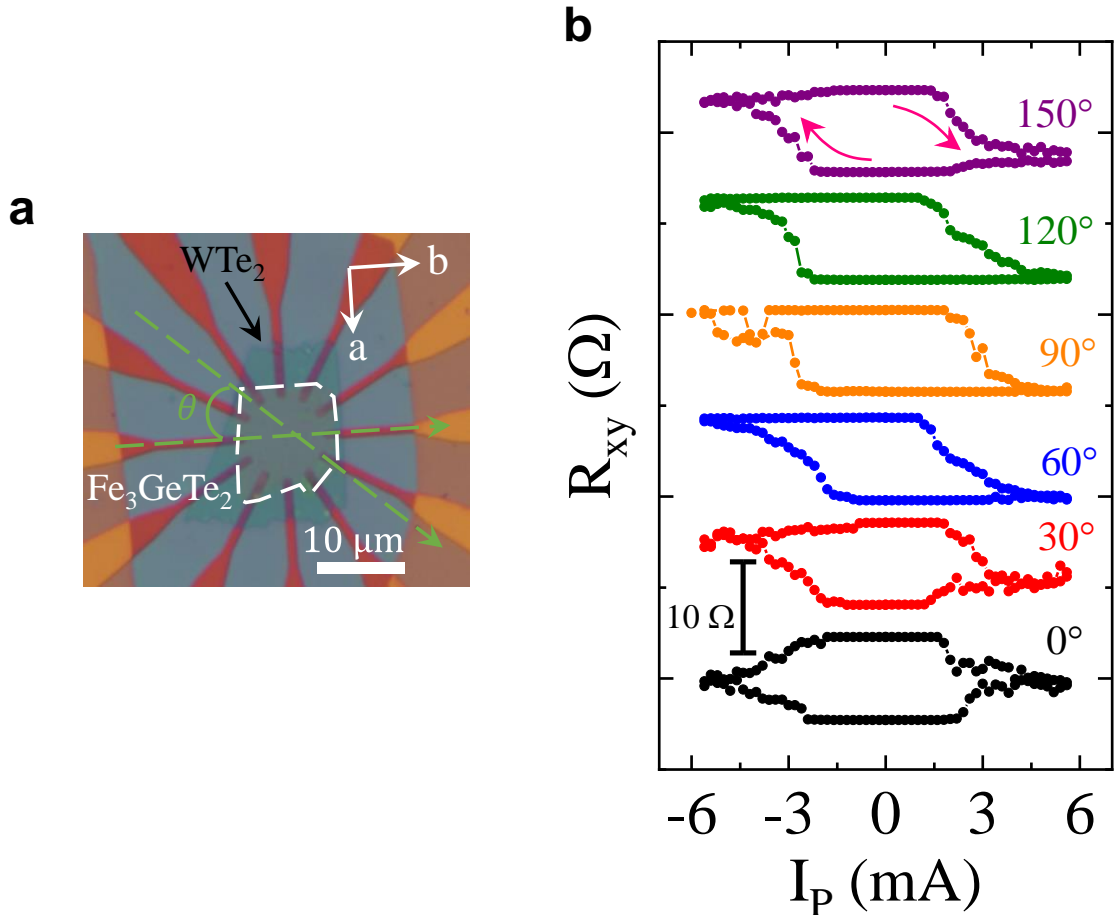

**Fig. S8 │ Reproducible results in device C. a,** The optical image of device C. **b,** The $R_{xy}$-$I_P$ loops at 120 K with $I_P$ applied along different angle $\theta$. Before sweeping $I_P$ from zero to large positive or negative values, the magnetization state is initialized by perpendicular magnetic field. The curves are shifted for clarity.

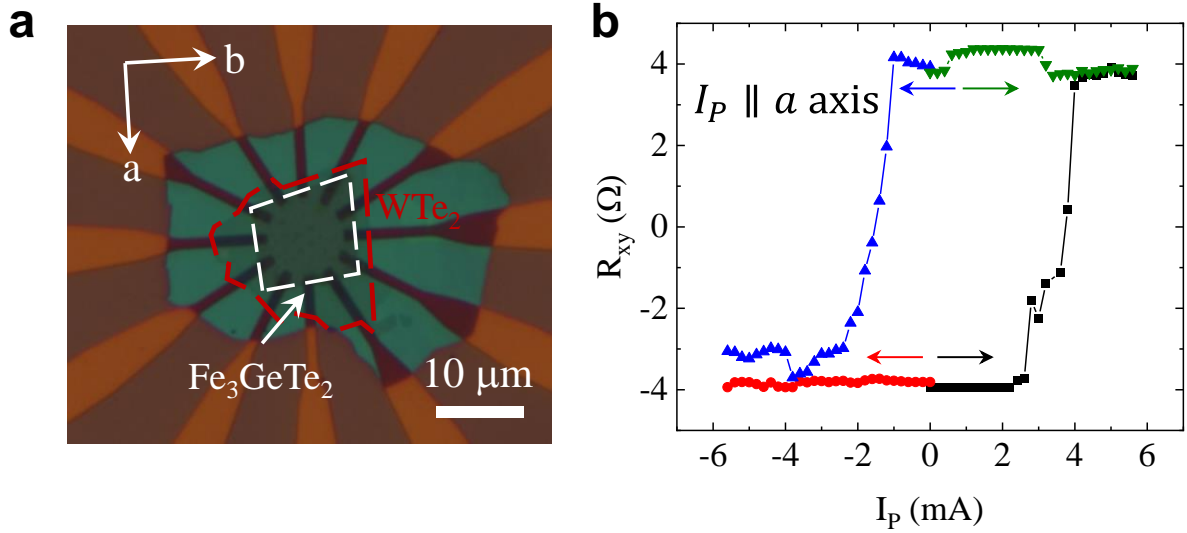

**Fig. S9 │ Reproducible results in device D. a,** The optical image of device D. **b,** Hall resistance as a function of pulse current I$_P$ at 120 K. I$_P$ is applied approximately along *a* axis. Before sweeping I$_P$ from zero to large positive or negative values, the magnetization state is initialized by applying perpendicular magnetic field.

**Table S1| Summary of the thickness of WTe$_2$ (t$_{WTe_2}$) and Fe$_3$GeTe$_2$ (t$_{FGT}$), and the switching current density $J_c$ at 120 K along *a* axis in different devices.**

| Device | $t_{WTe_2}$ (nm) | $t_{FGT}$ (nm) | $J_c$ ($10^{10} A/m^2$) |
|---|---|---|---|
| A | 11.9 | 11.2 | 8.5 |
| B | 14.7 | 12 | 8.6 |
| C | 5.6 | 14.4 | 6.5 |
| D | 3.5 | 12 | 8.4 |